\documentclass[aps,prl,twocolumn,tightenlines,superscriptaddress]{revtex4}
\usepackage{epsfig,rotating}
\usepackage{multirow}

\newcommand{\nuc}[2]{\hbox{$^{#1}$#2}}

\begin{document}
\title{In-beam $\gamma$-ray spectroscopy of \nuc{35}{Mg} and \nuc{33}{Na}} 

\author{A.\ Gade}
   \affiliation{National Superconducting Cyclotron Laboratory,
      Michigan State University, East Lansing, Michigan 48824, USA}
   \affiliation{Department of Physics and Astronomy,
      Michigan State University, East Lansing, Michigan 48824, USA}
\author{D.\ Bazin}
    \affiliation{National Superconducting Cyclotron Laboratory,
      Michigan State University, East Lansing, Michigan 48824, USA}
\author{B.\,A.\ Brown}
    \affiliation{National Superconducting Cyclotron Laboratory,
      Michigan State University, East Lansing, Michigan 48824, USA}
    \affiliation{Department of Physics and Astronomy,
      Michigan State University, East Lansing, Michigan 48824, USA}
\author{C.\,M.~Campbell}
    \affiliation{National Superconducting Cyclotron Laboratory,
      Michigan State University,
      East Lansing, Michigan 48824, USA}
    \affiliation{Department of Physics and Astronomy,
      Michigan State University, East Lansing, Michigan 48824, USA}
\author{J.\,M.\ Cook}
    \affiliation{National Superconducting Cyclotron Laboratory,
      Michigan State University, East Lansing, Michigan 48824, USA}
    \affiliation{Department of Physics and Astronomy,
      Michigan State University, East Lansing, Michigan 48824, USA}
\author{S.\ Ettenauer}
    \altaffiliation{Present Address: TRIUMF, 4004 Wesbrook Mall,
      Vancouver, B.C. V6T 2A3, Canada} 
    \affiliation{National Superconducting Cyclotron Laboratory,
      Michigan State University, East Lansing, Michigan 48824, USA}
\author{T.\ Glasmacher}
    \affiliation{National Superconducting Cyclotron Laboratory,
      Michigan State University, East Lansing, Michigan 48824, USA}
    \affiliation{Department of Physics and Astronomy,
      Michigan State University, East Lansing, Michigan 48824, USA}
\author{K.\,W.~Kemper}
   \affiliation{Department of Physics, Florida State University,
     Tallahassee, FL 32306, USA}
\author{S.\ McDaniel}
    \affiliation{National Superconducting Cyclotron Laboratory,
      Michigan State University, East Lansing, Michigan 48824, USA}
    \affiliation{Department of Physics and Astronomy,
      Michigan State University, East Lansing, Michigan 48824, USA}
\author{A.\ Obertelli}
    \affiliation{National Superconducting Cyclotron Laboratory,
      Michigan State University, East Lansing, Michigan 48824, USA}
\author{T.~Otsuka}
    \affiliation{Department of Physics and Center for Nuclear Study,
     University of Tokyo, Hongo, Tokyo 113-0033, Japan} 
    \affiliation{RIKEN, Hirosawa, Wako-shi, Saitama 351-0198, Japan}
\author{A.\ Ratkiewicz}
    \affiliation{National Superconducting Cyclotron Laboratory,
      Michigan State University, East Lansing, Michigan 48824, USA}
    \affiliation{Department of Physics and Astronomy,
      Michigan State University, East Lansing, Michigan 48824, USA}
\author{J.\ R.\ Terry}
    \affiliation{National Superconducting Cyclotron Laboratory,
      Michigan State University, East Lansing, Michigan 48824, USA}
    \affiliation{Department of Physics and Astronomy,
      Michigan State University, East Lansing, Michigan 48824, USA}
\author{Y.\ Utsuno}
    \affiliation{Japan Atomic Energy Agency, Tokai, Ibaraki 319-1195,
      Japan}
\author{D.\ Weisshaar}
    \affiliation{National Superconducting Cyclotron Laboratory,
      Michigan State University, East Lansing, Michigan 48824, USA}
\date{\today}

\begin{abstract}

Excited states in the very neutron-rich nuclei \nuc{35}{Mg} and
\nuc{33}{Na} were populated in the fragmentation of a \nuc{38}{Si}
projectile beam on a Be target at 83~MeV/u beam energy. We report on
the first observation of $\gamma$-ray transitions in \nuc{35}{Mg}, the
odd-$N$ neighbor of  
\nuc{34}{Mg} and \nuc{36}{Mg}, which are known to be part of the ``Island of
Inversion'' around $N=20$. The results are 
discussed in the framework of large-scale shell-model
calculations. For the $A=3Z$ nucleus \nuc{33}{Na}, a new 
$\gamma$-ray transition was observed that is suggested to complete the
$\gamma$-ray cascade $7/2^+ \rightarrow 5/2^+ \rightarrow 3/2^+_{gs}$
connecting three neutron 2p-2h intruder states that are predicted to
form a close-to-ideal $K=3/2$ 
rotational band in the strong-coupling limit.   

\end{abstract}

\maketitle

\section{Introduction}

The structure of nuclei in the so-called ``Island of
Inversion''~\cite{War90,Thi75}, a 
region of the nuclear chart comprised of neutron-rich Ne, Na and Mg
isotopes around neutron number $N=20$, has provided much insight into
the driving forces of shell evolution. In this region, driven largely by the
spin-isospin parts of the nuclear force~\cite{Ots01}, neutron
$n$p-$n$h ``intruder''
configurations of $\nu(sd)^{-n}(fp)^{+n}$ character (labeled
$n\hbar\omega$)
gain ``correlation energy''~\cite{Cau98} with respect to the
normal-order configurations and dominate the wave functions of low-lying states
including the ground states -- signaling the breakdown of
$N=20$ as a magic number in these nuclei. The resulting onset of
collectivity was 
quantified with inelastic scattering
experiments~\cite{Mot95,Pri99,Iwa01,Yan03,Chu05,Nie05} and
inferred from measurements of the $2^+_1$ excitation
energies~\cite{Det79,Yon01,Gad07,Doo09} on even-even nuclei
out to \nuc{32}{Ne} and \nuc{36}{Mg} at $Z=10$ and $Z=12$,
respectively. These collective properties of even-even nuclei are rather well
described by state-of-the-art Monte Carlo Shell-Model (MCSM)
calculations (SDPF-M effective interaction) that allow for
unrestricted mixing of neutron particle-hole configurations across the
$N=20$ shell gap. However, experimental information on their odd-$N$
neighbors,
\nuc{31,33}{Mg}~\cite{Num01,Pri02,Ney05,Yor07,Tri08,Mil09,Yor10,Tri10} 
and \nuc{30}{Na}~\cite{Ett08}, for example, has posed
significant challenges for these calculations and led to controversial
interpretations, indicating that odd-$A$ and odd-odd nuclei pose very
stringent benchmarks for a shell model description of this region.

In this paper we present results from the first $\gamma$-ray
spectroscopy of $^{35}_{12}$Mg$_{23}$ and report a new $\gamma$-ray
transition observed in the $A/Z=3$ nucleus \nuc{33}{Na}
($N=22$). Excited states in these very neutron-rich nuclei were
populated in the fragmentation of 
an intermediate-energy \nuc{38}{Si} rare-isotope beam on a \nuc{9}{Be}
target. The results are confronted with large-scale shell-model calculations.

\section{Experiment}

\begin{figure}[h]
        \epsfxsize 8.0cm
        \epsfbox{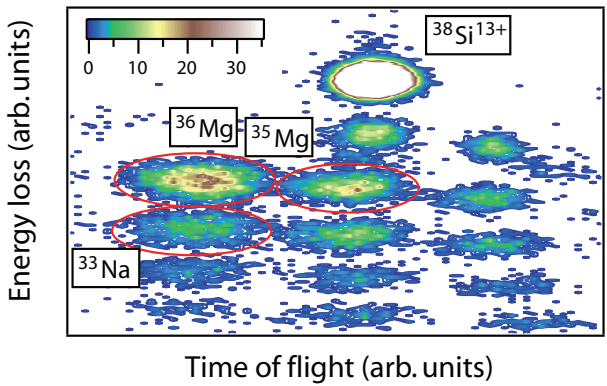}
\caption{\label{fig:pid} (Color online) Identification spectrum for the reaction
  residues produced in \nuc{9}{Be}(\nuc{38}{Si},\nuc{A}{Z})X at 
  83~MeV/u mid-target energy. All reaction residues can be
  unambiguously identified by plotting the energy loss measured in the
  S800 ionization chamber versus the ion's time of flight. The
  magnetic rigidity of the S800 spectrograph was set to center \nuc{36}{Mg}
  in the focal plane (see~\cite{Gad07}). The H-like charge
  state of the projectile beam - produced by electron
  pickup of the originally fully-stripped \nuc{38}{Si} passing through
  the reaction target - is the most intense constituent
  of the reacted beam.}
\end{figure}

The \nuc{38}{Si} secondary beam was produced by
fragmentation of 140~MeV/nucleon \nuc{48}{Ca} primary beam delivered by
the Coupled Cyclotron Facility of the National Superconducting Cyclotron
Laboratory onto a 752~mg/cm$^2$ primary \nuc{9}{Be} fragmentation target.
The isotope of interest was selected in the A1900 fragment
separator~\cite{a1900}, using a 750~mg/cm$^2$ Al wedge degrader for
purification. The \nuc{38}{Si} secondary beam interacted
with a 376(4)~mg/cm$^2$ thick \nuc{9}{Be} reaction target positioned
at the pivot point of the large-acceptance S800 spectrograph
~\cite{s800}. The reaction residues were identified on an
event-by-event basis from the time of flight taken between two
scintillators and the energy loss measured in the S800 ionization
chamber. The flight times were corrected for the reaction residue's
trajectories as reconstructed from the position and angle information
provided by the cathode readout drift chambers of the S800 focal-plane
detector system. The particle identification spectrum for the reaction
residues produced in  \nuc{9}{Be}(\nuc{38}{Si},\nuc{A}{Z})X is shown
in Fig.~\ref{fig:pid}.  

Since the measurement's main focus was the study of \nuc{36}{Mg} in
the two-proton knockout from \nuc{38}{Si}~\cite{Gad07,Sim09a,Sim09b},
the magnetic 
rigidity of the S800 spectrograph was set to center the longitudinal
momentum distribution of \nuc{36}{Mg} in the S800 focal
plane. Therefore, \nuc{35}{Mg} and \nuc{33}{Na} entered the focal
plane off-center and were impacted by acceptance cuts (estimated to
be of the order of 20-30\%). The cross sections for the production of
these nuclei from \nuc{38}{Si} were measured to be $\sigma
(\nuc{36}{Mg})=0.10(1)$~mb~\cite{Gad07}, $\sigma(\nuc{35}{Mg})\gtrsim
0.05$~mb and $\sigma(\nuc{33}{Na}) \gtrsim 0.03$~mb.       

\begin{figure}[h]
        \epsfxsize 7.5cm
        \epsfbox{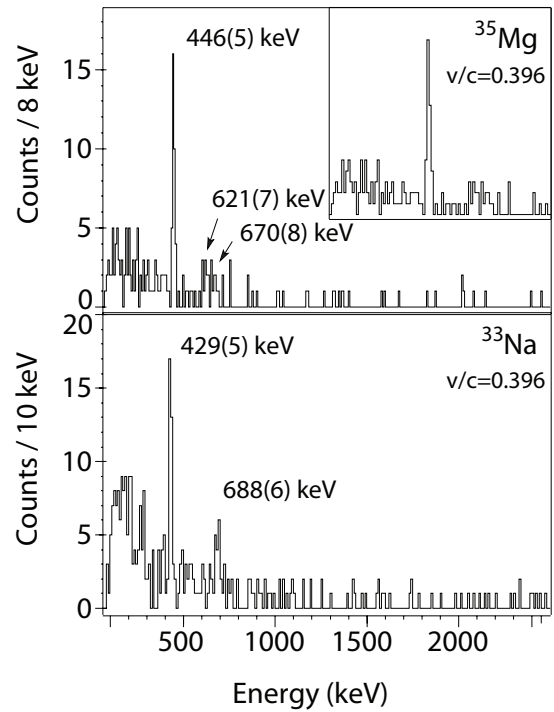}
\caption{\label{fig:gamma} Upper panel: Doppler-reconstructed $\gamma$-ray
  spectrum in coincidence with \nuc{35}{Mg} reaction residues.  Lower panel:
  Gamma-ray spectrum in coincidence with
  \nuc{33}{Na}.}
\end{figure}

The secondary \nuc{9}{Be} target was surrounded by
SeGA, an array of 32-fold segmented high-purity germanium 
detectors~\cite{sega}. The high degree of segmentation was used to
event-by-event Doppler reconstruct the $\gamma$ rays emitted by the
reaction residues in flight. For this, the location of the segment
that registered the largest energy deposition determines the
$\gamma$-ray emission angle relative to the target position. Sixteen
detectors were arranged in two rings (at 90$^\circ$ and 37$^\circ$ central
angles with respect to the beam axis). The 37$^\circ$ ring was
equipped with seven detectors while nine detectors occupied positions at
90$^\circ$. The array was calibrated for energy and efficiency with
\nuc{152}{Eu} and \nuc{226}{Ra} calibration standards. 

\section{Results}

Figure \ref{fig:gamma} shows the Doppler-reconstructed $\gamma$-ray spectra
measured in coincidence with \nuc{35}{Mg} and \nuc{33}{Na},
respectively. In the following sections we discuss the experimental
results in comparison to Monte-Carlo Shell-Model calculations using
the SDPF-M effective interaction~\cite{Uts99} that allows for
unrestricted mixing of neutron particle-hole configurations across the 
$N=20$ gap and to large-scale shell-model
calculations with the SDPF-U effective interaction~\cite{Now09} that
does not include neutron intruder configurations in the model space.  

\subsection{$^{35}$Mg}

For \nuc{35}{Mg}, a $\gamma$-ray transition at 446 keV
is clearly visible and possibly indications of two other transitions at 621 and
670~keV. There is no evidence for $\gamma$-ray peaks at higher
energies; in fact, the spectrum has remarkably few counts beyond
700~keV, likely indicative of a low neutron separation energy
( $S_n(\nuc{35}{Mg})=1020(200)$~keV reported in~\cite{Jur07} and
$S_n(\nuc{35}{Mg})=730(460)$~keV estimated in the compilation by 
G.\ Audi {\it et al.}~\cite{AW2003}).            

\begin{figure}[h]
        \epsfxsize 8.0cm
        \epsfbox{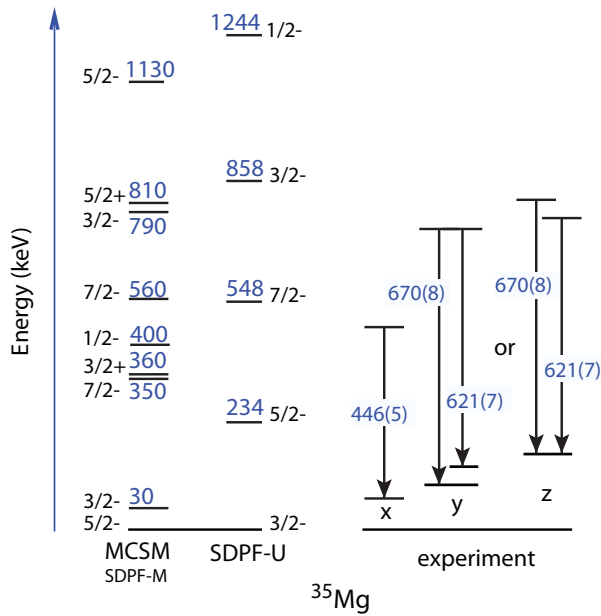}
\caption{\label{fig:compMg} (Color online) Right: Excited states below 1.2~MeV
  excitation energy predicted by the two shell-model
  calculations. Left: Possible arrangement of the experimentally
  observed transitions in the decay scheme of \nuc{35}{Mg}. A
  low-lying excited state below 80~keV could not have been detected
  with the threshold setting of the SeGA array.    }
\end{figure}

With its even-even neighbors \nuc{34}{Mg} and \nuc{36}{Mg} shown to be
part of the ``Island of Inversion'', one expects neutron 
particle-hole intruder excitations to dominate the structure of
\nuc{35}{Mg} as well. The Monte-Carlo Shell-Model calculation with the
SDPF-M effective 
interaction predicts eight excited states below 1.2~MeV, with a $5/2^-$
ground state almost degenerate with the first $3/2^-$ level at
30~keV excitation energy (see Figure~\ref{fig:compMg} and
Table~\ref{tab:mcsm_conf}). The 446~keV transition observed in the
experiment could correspond to the decay of any of the excited
$7/2^-_1$, $3/2^+$, $1/2^-$ or $7/2^-_2$ states between 350~keV and 560~keV
to either the ground state or the low-lying excited state predicted by
theory. We note that it was impossible for the measurement presented
here to detect a $\gamma$-ray transition below $\sim
80$~keV due to the energy threshold settings of the SeGA electronics
for this run. Assuming that the neutron separation energy is indeed
low for \nuc{35}{Mg}, the possible transitions at 621~keV and 670~keV
likely proceed to one or both of the $5/2^{-}_{1}$ or $3/2^{-}_{1}$
near-degenerate 
states. It might be possible that the two transitions originate from
the same level and that their energy difference of 49(11)~keV
corresponds to the energy spacing of the alleged $3/2^-$--$5/2^-$
doublet near the ground state. The possible scenarios of a decay level
scheme are summarized in Fig.~\ref{fig:compMg}, assuming that the
tentatively proposed 621 and 670~keV transitions are not feeding each
other based on the assumption of a low neutron separation energy. 

The reaction process leading to \nuc{35}{Mg} from \nuc{38}{Si} will
not predominantly proceed as the {\it direct} removal of two protons
and a neutron from \nuc{38}{Si} but is likely dominated by
the two-proton knockout into the continuum of \nuc{36}{Mg} and subsequent
neutron emission. Therefore one cannot expect the final-state and
nuclear structure selectivity of a strictly direct reaction and would rather
expect all bound low-lying states to be populated without preference
for certain configurations. This consideration together with the
predicted high level density in the MCSM and the exceptionally clean
$\gamma$-ray 
spectrum, showing evidence for at most three $\gamma$-ray transitions,
suggests that the neutron separation energy of \nuc{35}{Mg} is indeed
low -- likely at the lower end of $S_n=1020(200)$~keV
-- and that the $\gamma$-ray transitions we observe are not in 
coincidence and depopulate excited states at $\approx$500~keV and
$\approx$700~keV to the ground state or the alleged near-degenerate
excited state below 100~keV excitation energy (this corresponds to the
scenario in 
Fig.~\ref{fig:compMg} with x$\leq$80~keV and y=0~keV or z$\leq
80$~keV). 

\begin{figure}[h]
        \epsfxsize 8.0cm
        \epsfbox{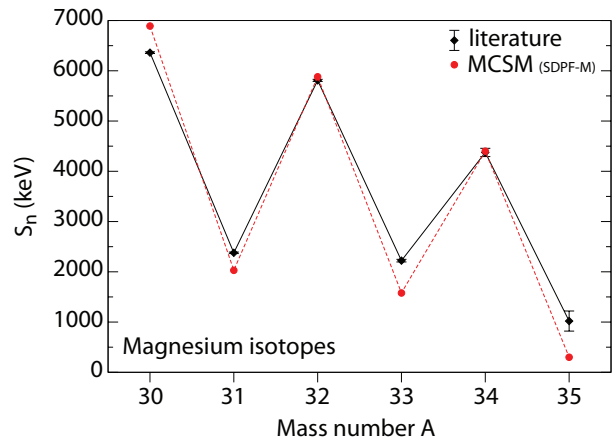}
\caption{\label{fig:separation} (Color online) Neutron separation
  energies, $S_n$, 
  for neutron-rich magnesium isotopes from literature ($A=30-33$ taken
  from~\cite{AW2003}, $A=34,35$ based 
  on the mass excess reported in~\cite{Jur07}) compared to the MCSM
  (SDPF-M) calculations. The overestimated odd-even staggering
  supports the assumption that pairing effects are overestimated,
  leading to an underestimation of $S_n$ for the odd-$A$ isotopes by
  several hundred keV. }
\end{figure}

      The
neutron-separation energy predicted in the MCSM calculation is too low
with $S_n \approx 300$~keV. This is likely related to the pairing matrix
elements in the interaction as evidenced by an overestimation of the
odd-even staggering in the calculation of $S_n$ in the magnesium
isotopes (see Fig.~\ref{fig:separation}).    

Table~\ref{tab:mcsm_conf} gives the $n\hbar\omega$ decomposition of
the wave functions of all negative and positive-parity states at
energies $E_x \leq 1.29$~MeV calculated with the MCSM. We note that
the first state, with 
$J^{\pi}=3/2^-$, that is dominated (56\%) by 0$\hbar\omega$ (non-intruder)
configurations occurs at 790~keV excitation energy. All other states
calculated in the MCSM in this energy window are almost pure intruder
states with $1\hbar\omega$ (positive-parity states) or $2\hbar\omega$
(negative-parity states) neutron intruder configurations.

For comparison, the conventional shell-model calculations with the
SDPF-U effective interaction predict only four excited states with
$J^{\pi}=5/2^-$, $7/2^-$, $3/2^-$ and $1/2^-$ on top of a $3/2^-$
ground state.

\begin{table}[h]
\begin{center}
 \vspace{0.5cm}
\caption{Composition of the \nuc{35}{Mg} wave functions with respect
  to $n \hbar\omega$ probabilities calculated within the MCSM. Only
  excited states with $E_x \leq 1.29$~MeV are listed. }
\begin{ruledtabular}
\begin{tabular}{ccccc}
$J^{+}$ & $E$(MeV)& 1$\hbar \omega$&3$\hbar\omega$&5$\hbar\omega$\\
              &    & (\%) & (\%) & (\%) \\
\hline
$3/2^+$ &  0.36 & 96.5 & 3.5 & 0.0\\
$5/2^+$ &  0.81 & 97.1 & 2.9 & 0.0\\
$7/2^+$ &  1.29 & 97.4 & 2.5 & 0.0\\ 
\hline
$J^{-}$ & $E$(MeV)& 0$\hbar \omega$&2$\hbar\omega$&4$\hbar\omega$\\
              &    & (\%) & (\%) & (\%) \\
\hline
$1/2^-$ &  0.40 & 1.7 & 97.8 & 0.4\\
$3/2^-$ &  0.03 & 6.9 & 92.5 & 0.6\\
        &  0.79 & 56.1 & 43.5 & 0.4\\
$5/2^-$ &  0.00 & 4.6 & 94.9 & 0.5\\
        &  1.13 & 10.1 & 89.6 & 0.4\\
$7/2^-$ &  0.35 & 12.4 & 87.0 & 0.6\\
        &  0.56 & 4.7 & 94.6 & 0.7\\
$9/2^-$ &  1.22 & 7.8 & 91.8 & 0.4
\label{tab:mcsm_conf}
\end{tabular}
\end{ruledtabular}
\end{center}
\end{table}

\subsection{ $^{33}$Na}

\begin{figure}[h]
        \epsfxsize 7.0cm
        \epsfbox{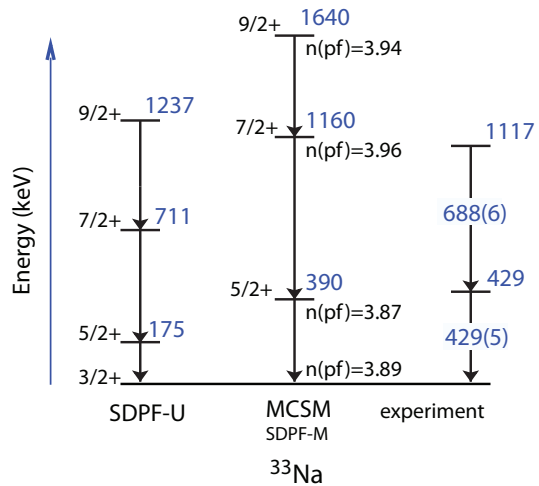}
\caption{\label{fig:compNa} (Color online) Level schemes of \nuc{33}{Na} as
  calculated in the shell model (left) compared to the measured
  $\gamma$-ray transitions. Assuming that the observed 429~keV and 688~keV
  transitions are in coincidence, the experimental level scheme agrees
  very well with the MCSM using the SDPF-M effective interaction. }
\end{figure}

The $\gamma$-ray spectrum taken in coincidence with \nuc{33}{Na} shows
two clear full-energy peaks that correspond to $\gamma$-ray
transitions at 429(5)~keV and 688(6)~keV in the most neutron-rich Na
isotope $(A=3Z)$ studied with $\gamma$-ray spectroscopy to date. A pioneering
experiment at RIBF in RIKEN populated one excited state at 467(13)~keV in
\nuc{33}{Na} with C-induced inelastic scattering and one-neutron
removal~\cite{Doo10}. We assume that this is the same $\gamma$-ray transition
observed by us, noting that the authors of~\cite{Doo10} obtained their
value from 
combining two very different energies of 476(12)~keV and 447(13)~keV
originating from their two different measurements, with the lower energy
closer to the value reported energy by us. In agreement
with~\cite{Doo10} and the systematics presented therein we propose
the 429(6)~keV $\gamma$-ray transition to proceed from the first
excited state to the ground state. The MCSM calculations with the
SDPF-M effective interaction predict the first excited state in
\nuc{33}{Na}, with $J^{\pi}=5/2^+$, at 390~keV on top of a $3/2^+$
ground state in good agreement with the measurement (see
Fig.~\ref{fig:compNa}). The calculated second
excited state, with $J^{\pi}=7/2^+$ at
$E_x=1.16$~MeV, is in good agreement with the data as well when
assuming that the $\gamma$-ray transition measured at 688~keV
depopulates the $7/2^+$ state and feeds the first excited $5/2^+$
state. In the MCSM, the neutron occupation 
numbers for the $pf$ shell, given in Fig.~\ref{fig:compNa}, are
close to $n(pf)=4$, indicating that all states shown in the calculated level
scheme are intruder states with 2$\hbar\omega$ neutron configurations
relative to two neutrons occupying the $pf$ shell in normal-order
filling. 
                
From the systematic studies of the onset of intruder configurations
and quadrupole collectivity in the chain of Na isotopes approaching
$N=20$~\cite{Utsuno04}, the question arises how the quadrupole
collectivity and deformation 
evolves for more neutron-rich isotopes beyond $N=20$. For this, the
quadrupole moments and $B(E2)$ transition strengths for the $3/2^+$,
$5/3^+$, and $7/2^+$ states in \nuc{33}{Na} were calculated with the
MCSM using $e_p=1.3e$ and $e_n=0.5e$ effective charges
(Table~\ref{tab:collective}). From the experimental transition
energies and the calculated $B(E2;7/2^+ \rightarrow
3/2^+)=67.5$~e$^2$fm$^4$, $B(E2; 7/2^+ \rightarrow
5/2^+)=99.75$~e$^2$fm$^4$  values together with the predicted rather  
strong $B(M1;7/2^+ \rightarrow 5/2^+)=0.59$~$\mu_N^2$ reduced M1
transition probability, the $\gamma$-ray branch for the 
decay to the ground state is estimated to be only 4.2\% of the
$\gamma$-decay branch to the $5/2^+$ state, impossible to be observed
in the experiment at
our level of statistics.  Assuming that these states form a 
rotational band with $K=3/2$ in the strong-coupling limit, the
intrinsic quadrupole moments $Q_0$ were deduced. As shown in
Table~\ref{tab:collective}, all quadrupole moments and $B(E2)$ values
are well described with a common $Q_0 \sim 70$~fm$^2$. This indicates
that in the MCSM calculation this structure is close to the ideal
case of a well-deformed $K=3/2$ rotational band in the strong coupling
limit. This is further supported by the excitation energies; the ratio
of the energy differences $E(7/2^+)-E(3/2^+)$ and $E(5/2^+)-E(3/2^+)$
in an ideal $K=3/2$ rotational band is expected to be 2.4, while the
measured energies in \nuc{33}{Na} give a very similar ratio of
2.6. For comparison, the same energy ratio in
\nuc{31}{Na}~\cite{Doo10} gives 3.1, indicating that the low-lying
states in Na at $N=22$ are closer to an ideal $K=3/2$ rotational band
than at $N=20$. 

To put the predicted deformation into perspective, for the $N=22$
isotone \nuc{34}{Mg}, the MCSM calculates $B(E2; 0^+ \rightarrow
2^+_1)=552$~e$^2$fm$^4$ (in agreement with
experiment~\cite{Iwa01,Chu05}) and $Q(2^+_1)=-21.4$~fm$^2$, yielding
an intrinsic quadrupole moment of $Q_0=75$~fm$^2$, very similar to
\nuc{33}{Na}. At $N=20$, \nuc{32}{Mg}, the
calculated $B(E2;0^+ \rightarrow 2^+_1)=447$~e$^2$fm$^4$ (in agreement
with experiment \cite{Mot95,Pri99,Iwa01,Chu05}) and
$Q(2^+_1)=-17.1$~fm$^2$ lead to $|Q_o|=67$~fm$^2$ and 60~fm$^2$,
respectively, with the small discrepancy potentially indicative of
triaxiality or gamma-softness.    

\begin{table}[h]
\begin{center}
 \vspace{0.5cm}
\caption{Quadrupole moments, $E2$ transition strengths and extracted
  intrinsic quadrupole moments $Q_0$ for the
  $3/2^+$, $5/2^+$, and $7/2^+$ lowest-lying states in
  \nuc{33}{Na} calculated with the MCSM (SDPF-M with proton and neutron
  effective charges of $e_p=1.3e$ and $e_n=0.5e$). $Q_0$ was deduced
  assuming that this is a $K=3/2$ rotational band within the
  strong-coupling limit. With this assumption, the quadrupole
  moments and transition strengths in this band are very well
  described with a common intrinsic quadrupole moment of $Q_0 \sim 70$~fm$^2$,
  indicating that this structure in \nuc{33}{Na} is indeed a
  rather well-deformed $K=3/2$ rotational band within the MCSM calculation. }
\begin{ruledtabular}
\begin{tabular}{ccc}
$J$ & $Q$ & $Q_0$\\
        & (fm$^2$)  & (fm$^2$)  \\
\hline
$3/2^+$ &  +14.3 & 72 \\
$5/2^+$ &  -4.3 & 60 \\
$7/2^+$ &  -14.6 & 73 \\ 
\hline
$J_i \rightarrow J_f$ & $B(E2)$     & $|Q_o|$\\
                      & (e$^2$fm$^4$)& (fm$^2$) \\
\hline
$3/2^+ \rightarrow 5/2^+$ & 263 & 72\\
$3/2^+ \rightarrow 7/2^+$ & 135 & 69\\
$5/2^+ \rightarrow 7/2^+$ & 133 & 68
\label{tab:collective}
\end{tabular}
\end{ruledtabular}
\end{center}
\end{table}

The level scheme predicted by the conventional shell model
calculations based on the SDPF-U effective interaction, which does not
allow for neutron intruder configurations, has the levels in
the same order but predicts the $5/2^+$ first excited state to be within
175~keV of the $3/2^+$ ground state (see Fig.~\ref{fig:compNa}). A
strong $\gamma$-ray transition above 90-100~keV would have been
observed in the present measurement.  

\section{Summary}
In summary, in-beam $\gamma$-ray spectroscopy of the very neutron-rich
nuclei \nuc{35}{Mg} and
\nuc{33}{Na} was performed following the fragmentation of a
\nuc{38}{Si} projectile beam on a \nuc{9}{Be} target at intermediate beam
energy. In \nuc{35}{Mg}, the odd-$N$ neighbor of
the ``Island of Inversion'' 
nuclei \nuc{34}{Mg} and \nuc{36}{Mg}, $\gamma$-ray transitions were
measured for the first time. In comparison to Monte-Carlo Shell-Model
calculations with the 
SDPF-M effective interaction, the transitions are interpreted as
connecting excited states around $\sim 450$~keV and $\sim 700$~keV
to the $5/2^-$ ground state and/or the first excited $3/2^-$ state
that is predicted to be within 30~keV of the ground state. For the
$N=2Z$ nucleus \nuc{33}{Na}, the most neutron-rich Na isotope studied
to date with $\gamma$-ray spectroscopy, a new $\gamma$-ray transition at
688(6)~keV was measured in addition to the known transition at
429(5)~keV. The two transitions, if in coincidence, are in very good
agreement with MCSM calculations and proposed to establish the $7/2^+
\rightarrow 5/2^+ \rightarrow 3/2^+_{gs}$ cascade of the ground-state
band in \nuc{33}{Na} -- predicted in the MCSM to be an almost ideal
$K=3/2$ rotational band structure (of intruder states) in the 
strong-couping limit. Coulomb excitation or excited-state lifetime
measurements are needed to confirm the degree of deformation and
rotational character and remain a challenge for future experiments. All
low-lying states in these two nuclei are predicted to be intruder
states, putting \nuc{35}{Mg} and \nuc{33}{Na} inside the ``Island of
Inversion''.

\begin{acknowledgments}
This work was supported by the National Science Foundation under
Grants No. PHY-0606007 and PHY-0758099 and in part by the JSPS
Core-to-Core Program EFES and by a Grant-in-Aid for Young Scientists
(No. 21740204) from the MEXT of Japan and by a Grant-in-Aid
for Specially Promoted Research (No. 13002001) from the MEXT.
\end{acknowledgments}


\begin{thebibliography}{10}
\bibitem{War90} E.\ K.\ Warburton, J.\ A.\ Becker and B.\ A.\ Brown,
  Phys.\ Rev.\ C {\bf 41}, 1147 (1990). 
\bibitem{Thi75} C.\ Thibault {\it et al.}, Phys.\ Rev.\ C {\bf 12}, 644
  (1975).
\bibitem{Ots01} T.\ Otsuka, R.\ Fujimoto, Y.\ Utsuno,
  B.\ A.\  Brown, M.\ Honma, and T.\ Mizusaki, Phys.\ Rev.\ Lett.\ {\bf 87},
  082502 (2001).
\bibitem{Cau98} E.\ Caurier, F.\ Nowacki, A.\ Poves, and J.\ Retamosa,
Phys. Rev. C {\bf 58}, 2033 (1998).  

\bibitem{Mot95} T.\ Motobayashi {\it et al.}, Phys.\ Lett.\ B {\bf
  346}, 9 (1995). 
\bibitem{Pri99} B.\ V.\ Pritychenko {\it et al.}, Phys.\ Lett.\ B {\bf
  461}, 322 (1999).

\bibitem{Iwa01} H.\ Iwasaki {\it et al.}, Phys.\ Lett.\ B {\bf 522},
  227 (2001).
\bibitem{Yan03} Y.\ Yanagisawa {\it et al.}, Phys.\ Lett.\ B {\bf
  566}, 84 (2003). 
\bibitem{Chu05} J.\ A.\ Church, C.\ M.\ Campbell, D.-C.\ Dinca, J.\ Enders,
  A.\ Gade, T.\ Glasmacher, Z.\ Hu, R.\ V.\ F.\ Janssens,
  W.\ F.\ Mueller, H.\ Olliver,
  B.\ C.\ Perry, L.\ A.\ Riley, K.\ L.\ Yurkewicz, Phys.\ Rev.\ C {\bf 72}, 054320
  (2005). 
\bibitem{Nie05} O.\ Niedermaier {\it et al.}, Phys.\ Rev.\ Lett.\ {\bf
  94}, 172501 (2005).
\bibitem{Det79} C.\ D{\'e}traz, D.\ Guillemaud, G.\ Huber,
  R.\ Klapisch, M.\ Langevin, F.\ Naulin, C.\ Thibault, L.\ C.\ Carraz, and
  F.\ Touchard, Phys.\ Rev.\ C {\bf 19}, 164
  (1979).
\bibitem{Yon01} K.\ Yoneda {\it et al.}, Phys.\ Lett.\ B {\bf 499},
  233 (2001).
\bibitem{Gad07} A.\ Gade, P.\ Adrich, D.\ Bazin, M.\ D.\ Bowen,
  B.\ A.\ Brown, C.\ M.\ Campbell, J.\ M.\ Cook, S.\ Ettenauer,
  T.\ Glasmacher, K.\ W.\ Kemper, S.\ McDaniel, A.\ Obertelli, T.\ Otsuka,
  A.\ Ratkiewicz, K.\ Siwek, J.\ R.\ Terry, J.\ A.\ Tostevin, Y.\ Utsuno, and
  D.\ Weisshaar, Phys.\ Rev.\ Lett.\ {\bf 99}, 072502 (2007).
\bibitem{Doo09} P.\ Doornenbal {\it et al.}, Phys.\ Rev.\ Lett.\ {\bf 103},
  032501 (2009). 
\bibitem{Num01} S.\ Nummela {\it et al.}, Phys.\ Rev.\ C {\bf 64}, 054313
  (2001). 
\bibitem{Pri02} B. V. Pritychenko, T. Glasmacher, P. D. Cottle,
  R. W. Ibbotson,  K. W. Kemper, L. A. Riley, A. Sakharuk, H. Scheit,
  M. Steiner, and V. Zelevinsky, Phys. Rev. C {\bf 65}, 061304 (2002). 
\bibitem{Ney05} G. Neyens, M. Kowalska, D. Yordanov, K. Blaum,
  P. Himpe, P. Lievens, S. Mallion, R. Neugart, N. Vermeulen,
  Y. Utsuno, and T. Otsuka, Phys.\ Rev.\ Lett.\ {\bf 94},
  022501 (2005).
\bibitem{Yor07}  D. T. Yordanov, M. Kowalska, K. Blaum, M. De Rydt,
  K. T. Flanagan, P. Lievens, R. Neugart, G. Neyens, and H. H. Stroke,
  Phys. Rev. Lett. {\bf 99}, 212501 (2007) 
\bibitem{Tri08} Vandana Tripathi, S. L. Tabor, P. F. Mantica,
  Y. Utsuno, P. Bender, J. Cook, C. R. Hoffman, Sangjin Lee,
  T. Otsuka, J. Pereira, M. Perry, K. Pepper, J. S. Pinter, J. Stoker,
  A. Volya, and D. Weisshaar, Phys. Rev. Lett. {\bf 101}, 142504 (2008). 
\bibitem{Mil09} D.\ Miller, P.\ Adrich, B.\ A.\ Brown, V.\ Moeller,
  A.\ Ratkiewicz, W.\ Rother, K.\ Starosta, J.\ A.\ Tostevin,
  C.\ Vaman, and P.\ Voss, Phys.\ Rev.\ C {\bf 79}, 054306 (2009).
\bibitem{Yor10} D. T. Yordanov, K. Blaum, M. De Rydt, M. Kowalska,
   R. Neugart, G. Neyens, and I. Hamamoto, Phys. Rev. Lett. {\bf 104},
   129201 (2010). 
\bibitem{Tri10} Vandana Tripathi, S. L. Tabor, P. F. Mantica,
  Y. Utsuno, P. Bender, J. Cook, C. R. Hoffman, Sangjin Lee,
  T. Otsuka, J. Pereira, M. Perry, K. Pepper, J. S. Pinter, J. Stoker,
  A. Volya, and D. Weisshaar, Phys. Rev. Lett. {\bf 104}, 129202 (2010).  
\bibitem{Ett08} S. Ettenauer, H. Zwahlen, P. Adrich, D. Bazin,
  C. M. Campbell, J. M. Cook, A. D. Davies, D.-C. Dinca, A. Gade,
  T. Glasmacher, J.-L. Lecouey, W. F. Mueller, T. Otsuka,
  R. R. Reynolds, L. A. Riley, J. R. Terry, Y. Utsuno, and K. Yoneda,
  Phys.\ Rev.\ C {\bf 78}, 017302 (2008).
\bibitem{Uts99} Y. Utsuno, T.\ Otsuka, T.\ Mizusaki, M.\ Honma,
  Phys. Rev. C {\bf 60}, 054315 (1999). 
\bibitem{a1900} D.\ J.\ Morrissey {\it et al.}, Nucl.\ Instrum.\ Methods
  in Phys.\ Res.\ B {\bf 204}, 90 (2003).
\bibitem{s800} D.\ Bazin {\it et al.}, Nucl.\ Instrum.\ Methods in Phys.\
  Res.\ B {\bf 204}, 629 (2003).
\bibitem{Sim09a} E.C. Simpson, J.A. Tostevin, D. Bazin, B.A. Brown, and
  A. Gade, Phys.\ Rev.\ Lett.\ {\bf 102}, 132502 (2009).
\bibitem{Sim09b} E.C. Simpson, J.A. Tostevin, D. Bazin, and
  A. Gade, Phys.\ Rev.\ C {\bf 79}, 064621 (2009).
\bibitem{sega}  W.\ F.\ Mueller {\it et al.}, Nucl.\ Instr.\ and
  Meth.\  A {\bf 466}, 492 (2001).
\bibitem{Now09} F.\ Nowacki and A.\ Poves, Phys.\ Rev.\ C {\bf 79},
  014310 (2009). 
\bibitem{Jur07} B.\ Jurado, H.\ Savajols, W.\ Mittig, N.\ A.\ Orr,
  P.\ Roussel-Chomaz, D.\ Baiborodin, W.\ N.\ Catford, M.\ Chartier,
  C.\ E.\ Demonchy, Z.\ Dlouhy, A.\ Gillibert, L.\ Giot, A.\ Khouaja,
  A.\ Lepine-Szily, S.\ Lukyanov, J.\ Mrazek, Y.\ E.\ Penionzhkevich,
  S.\ Pita, M.\ Rousseau, A.\ C.\ Villari, Phys.\ Lett.\ B {\bf 649}, 43 (2007). \bibitem{AW2003} G.\ Audi, A.\ H.\ Wapstra, and C.\ Thibault,
  Nucl.\ Phys.\ {\bf A 729}, 337 (2003).

\bibitem{Doo10} P.\ Doornenbal {\it et al.}, Phys.\ Rev.\ C {\bf 81},
  041305 (2010).  
\bibitem{Utsuno04} Y.\ Utsuno, T. Otsuka, T. Glasmacher, T. Mizusaki,
  and M. Honma, Phys.\ Rev.\ C {\bf 70}, 044307 (2004). 

\end{thebibliography}
\end{document}